\documentstyle[12pt]{article}
\catcode`\@=11

\@addtoreset{equation}{section}
\def\theequation{\arabic{section}.\arabic{equation}}
\catcode`\@=11

\def\section{\@startsection{section}{1}{\z@}{3ex plus 1ex minus
   .2ex}{2ex plus .2ex}{\large\bf}}

\def\eqnarray{\let\@currentlabel=\theequation\refstepcounter{equation}
    \global\@eqnswtrue
    \global\@eqcnt\z@\tabskip\@centering\let\\=\@eqncr
    $$\halign to \displaywidth\bgroup\@eqnsel\hskip\@centering
      $\displaystyle\tabskip\z@{##}$&\global\@eqcnt\@ne
       \hfil${{}##{}}$\hfil
      &\global\@eqcnt\tw@ $\displaystyle\tabskip\z@{##}$\hfil
       \tabskip\@centering&\llap{##}\tabskip\z@\cr}
\def\lefteqn#1{\hbox to 4\arraycolsep{$\displaystyle #1$\hss}}
\def\thesection{\arabic{section}}

\def\appendix{\setcounter{section}{0}
        \def\thesection{Appendix \Alph{section}}
        \def\theequation{\Alph{section}.\arabic{equation}}}

\long\def\@makefntext#1{\parindent 0cm\noindent
\hbox to 1em{\hss$^{\@thefnmark}$}#1}
\def\rref#1{(\ref{#1})}
\def\Tr{{\mathit{Tr}}}
\newcommand{\beq}{\begin{equation}}
\newcommand{\eeq}{\end{equation}}
\begin{document}
%%%%%%%%%%%%%%%%%%%%%%%%%%%%%%%%%%%%%%%%%%%%%%%%%%%%%%%%%%%%%%%%%%%%%%%%%%%
%     C I T E . S T Y
%     compressed lists of numerical citations: [11-16]
%     see also OVERCITE.STY and DRFTCITE.STY
%
%     Copyright (C) 1989-1992 by Donald Arseneau
%     These macros may be freely transmitted, reproduced, or modified for
%     non-commercial purposes provided that this notice is left intact.
%
%
%  \@citen contains the code that parses the list of names, ignoring
%  spaces after commas, writes the aux file \citation, and formats the
%  number list.  \citen can be used by itself to give citation numbers
%  without the other formatting; e.g., "See also ref.~\citen{junk}."
%
\def\citen#1{%
\edef\@tempa{\@ignspaftercomma,#1, \@end, }% ignore spaces in parameter list
\edef\@tempa{\expandafter\@ignendcommas\@tempa\@end}%
\if@filesw \immediate \write \@auxout {\string \citation {\@tempa}}\fi
\@tempcntb\m@ne \let\@h@ld\relax \let\@citea\@empty
\@for \@citeb:=\@tempa\do {\@cmpresscites}%
\@h@ld}
%
% for ignoring spaces in the input:
\def\@ignspaftercomma#1, {\ifx\@end#1\@empty\else
   #1,\expandafter\@ignspaftercomma\fi}
\def\@ignendcommas,#1,\@end{#1}
%
% For each citation, check if it is defined, if it is a number, and
% if it is a consecutive number that can be represented like 3-7.
%
\def\@cmpresscites{%
 \expandafter\let \expandafter\@B@citeB \csname b@\@citeb \endcsname
 \ifx\@B@citeB\relax % undefined
    \@h@ld\@citea\@tempcntb\m@ne{\bf ?}%
    \@warning {Citation `\@citeb ' on page \thepage \space undefined}%
 \else%  defined
    \@tempcnta\@tempcntb \advance\@tempcnta\@ne
    \setbox\z@\hbox\bgroup % check if citation is a number:
    \ifnum\z@<0\@B@citeB \relax
       \egroup \@tempcntb\@B@citeB \relax
       \else \egroup \@tempcntb\m@ne \fi
    \ifnum\@tempcnta=\@tempcntb % Number follows previous--hold on to it
       \ifx\@h@ld\relax % first pair of successives
          \edef \@h@ld{\@citea\@B@citeB}%
       \else % compressible list of successives
%         % use \hbox to avoid easy \exhyphenpenalty breaks
          \edef\@h@ld{\hbox{--}\penalty\@highpenalty \@B@citeB}%
       \fi
    \else   %  non-successor--dump what's held and do this one
       \@h@ld \@citea \@B@citeB \let\@h@ld\relax
 \fi\fi%
 \let\@citea\@citepunct
}
%
%%    To put space after the comma, use:
\def\@citepunct{,\penalty\@highpenalty\hskip.13em plus.1em minus.1em}%
%%    For no space after comma, use:
%% \def\@citepunct{,\penalty\@highpenalty}%
%%
%
%  Make \@citex refer to \citen:
%
\def\@citex[#1]#2{\@cite{\citen{#2}}{#1}}%
%
%  Replacement for \@cite.  Give one normal space before the citation,
%  set high penalties for linebreaks,
%
\def\@cite#1#2{\leavevmode\unskip
  \ifnum\lastpenalty=\z@ \penalty\@highpenalty \fi % highpenalty before
  \ [{\multiply\@highpenalty 3 #1% % triple-highpenalties within list
      \if@tempswa,\penalty\@highpenalty\ #2\fi % and before note.
    }]\spacefactor\@m}
\let\nocitecount\relax  % in case \nocitecount was used for drftcite
%
%%%%%%%%%%%%%%%%%%%%%%%%%%%%%%%%%%%%%%%%%%%%%%%%%%%%%%%%%%%%%%%%%%%%%%%%%%
\begin{titlepage}
\vspace{.5in}
\begin{flushright}
UCD-98-9\\
June 1998\\
hep-th/9806026\\
\end{flushright}
\vspace{.5in}
\begin{center}
{\Large\bf
What We Don't Know\\[1.2ex]
about BTZ Black Hole Entropy}\\
\vspace{.4in}
{S.~C{\sc arlip}\footnote{\it email: carlip@dirac.ucdavis.edu}\\
       {\small\it Department of Physics}\\
       {\small\it University of California}\\
       {\small\it Davis, CA 95616}\\{\small\it USA}}
\end{center}

\vspace{.5in}
\begin{center}
\begin{minipage}{5in}
\begin{center}
{\large\bf Abstract}
\end{center}
{\small With the recent discovery that many aspects of black hole
thermodynamics can be effectively reduced to problems in three spacetime
dimensions, it has become increasingly important to understand the
``statistical mechanics'' of the (2+1)-dimensional black hole of
Ba{\~n}ados, Teitelboim, and Zanelli (BTZ).  Several conformal field
theoretic derivations of the BTZ entropy exist, but none is completely
satisfactory, and many questions remain open: there is no consensus
as to what fields provide the relevant degrees of freedom or where
these excitations live.  In this paper, I review some of the unresolved
problems and suggest avenues for their solution.
}
\end{minipage}
\end{center}
\end{titlepage}
\addtocounter{footnote}{-1}

\section{Introduction \label{sec1}}

Since its discovery in 1992, the (2+1)-dimensional black hole of
Ba{\~n}ados, Teitelboim, and Zanelli \cite{BTZ,BHTZ} has served as
a useful  model for realistic black hole physics \cite{Carlip1}.
Interest in this model has recently heightened with the discovery
that the thermodynamics of higher-dimensional black holes can often
be understood in terms of the BTZ solution.  Many of the black holes
relevant to string theory have near-horizon geometries of the form
$\hbox{\it BTZ}\times M$, where $M$ is a simple manifold, and their
entropies and grey-body factors can be obtained the BTZ black hole
directly \cite{Strominger,Birmingham,Balasubramanian,Iofa,Alwis,Kaloper,%
Maldacena,Lee,Martinec,Iofa2,Behrndt,Sachs,Teo2,Lee2,Cvetic,Cvetic2}
or by duality \cite{Hyun,Sfetsos,Teo}.  It has become vital to
understand BTZ black hole thermodynamics from first principles.

A natural guess is that black hole entropy should be associated
with horizon degrees of freedom.  In 2+1 dimensions, the horizon is
two-dimensional, suggesting the relevance of conformal field theory.
This connection is strengthened by the observation that (2+1)-dimensional
gravity can be written as a Chern-Simons theory \cite{Achucarro,Witten},
and that Chern-Simons theories induce Wess-Zumino-Witten models on
boundaries \cite{Witten2,EMSS}.

The first application of conformal field theory techniques to BTZ black
hole entropy appeared in Ref.\ \citen{Carlip2}, with the treatment
of boundary conditions later simplified in Ref.\ \citen{Banados1}.
This approach explicitly counts states, but relies on a poorly
understood nonunitary theory and a complicated treatment of boundary
data.  It was recently shown that the same technique gives the correct
entropy for (2+1)-dimensional de Sitter space \cite{Strominger2}.  A
simpler computation, based on the Euclidean partition function, was
developed in Ref.\ \citen{Carlip3}.  Like most partition function
methods, however, this approach does not explicitly display the states
being counted, and it involves an analytic continuation from Lorentzian
to Riemannian metrics that is not completely understood.  A rather
different Euclidean partition function approach appeared in Ref.\
\citen{BBO}.

Recently, Strominger has suggested a much simpler derivation of the BTZ
black hole entropy \cite{Strominger}.  He begins with the observation,
known since 1986 \cite{Brown}, that the asymptotic symmetry group of
(2+1)-dimensional gravity with a negative cosmological constant
$\Lambda=-1/\ell^2$ is generated by two copies of the Virasoro algebra,
with central charges
\beq
c_L = c_R = {3\ell\over2G} .
\label{a1}
\eeq
The degrees of freedom, now at infinity rather than the horizon, are
thus described by a conformal field theory with this central charge.
The asymptotic density of states for such a theory follows from a result
of Cardy's \cite{Cardy,Cardy2}: it is
\beq
\ln \rho(\Delta,{\bar\Delta}) \sim
  2\pi\sqrt{c_R\Delta\over6} + 2\pi\sqrt{c_L{\bar\Delta}\over6} ,
\label{a2}
\eeq
where $\Delta$ and $\bar\Delta$ are the eigenvalues of the two Virasoro
generators $L_0$ and ${\bar L}_0$.  But for the BTZ black hole, we have
(up to an ambiguous additive constant) \cite{Banados2}
\beq
M = {(L_0 + {\bar L}_0)/\ell} , \quad J = L_0 - {\bar L}_0  ,
\label{a3}
\eeq
where $r_\pm$ are the radii of the inner and outer horizons.  Substituting
into \rref{a2} and using the expression  \rref{A4} for the mass and angular
momentum, we obtain the correct entropy
\beq
S = {2\pi r_+\over4G} .
\label{a4}
\eeq

Attractive as this approach is, it is not yet the full answer.  The
Cardy formula \rref{a2} is derived from the partition function, and
like the Euclidean approach of Ref.\ \citen{Carlip3}, it hides the
actual degrees of freedom that contribute to the entropy.  Moreover, as
we shall see below, equation \rref{a2} involves some hidden assumptions
that may not hold for the BTZ black hole.  Strominger's derivation also
raises the question of where the relevant degrees of freedom are located:
other approaches describe excitations at the horizon, but the central
charges \rref{a1} are relevant for a conformal field theory at infinity.

In this paper, I will discuss these issues, describing some of the
assumptions and ambiguities in various approaches and suggesting
a few paths forward.  While I have tried to make this work reasonably
self-contained, I assume some familiarity with the BTZ solution (see
Ref.\ \citen{Carlip1} for a review).  An appendix describes some
coordinate systems and conventions, and a second appendix discusses
issues related to the choice of boundary conditions in Ref.\
\citen{Carlip2}.

It is perhaps worth emphasizing that although some of the approaches
described here are inspired by string theory, my focus is on pure
(2+1)-dimensional gravity.  A full string theoretical picture of
the BTZ black hole presumably involves a large number of added degrees
of freedom (see, for example, \cite{Giveon}), and the relationship
to the counting techniques described here is not entirely clear.

\section{Counting States: Partition Functions and Cardy's Formula
         \label{sec2}}

In the microcanonical ensemble, the entropy is essentially the logarithm
of the density of states $\rho(E)$.  There are two common methods for
determining this quantity.  The most straightforward, which I address
in section \ref{sec3}, is to simply count: we begin with a vacuum state
and see how many different ways we can add excitations to reach the
energy $E$.  The second approach is less direct, but often simpler: we
manipulate the partition function to obtain a density of states.  For
conformal field theories, this method yields the Cardy formula \rref{a2}
and its generalization, equation \rref{b14} below.

We begin with a conformal field theory with central charge $c$, with
the standard Virasoro algebra
\beq
[L_m,L_n] = (m-n)L_{m+n} + {c\over12}m(m^2-1)\delta_{m+n,0} .
\label{b2}
\eeq
Cardy's basic result \cite{Cardy,Cardy2} is that the quantity
\beq
Z_0(\tau,\bar\tau) =
\Tr\ e^{2\pi i(L_0-{c\over24})\tau}e^{-2\pi i({\bar L}_0-{c\over24}){\bar\tau}}
\label{b3}
\eeq
is modular invariant, and in particular invariant under the transformation
$\tau\rightarrow-1/\tau$.  This argument involves only some quite general
properties of conformal field theories, and I will assume it holds for
the theory associated with the BTZ black hole.

Now, the partition function on the torus of modulus $\tau$ is
\beq
Z(\tau,\bar\tau) =
\Tr\ e^{2\pi i\tau L_0}e^{-2\pi i{\bar\tau}{\bar L}_0} =
\sum\rho(\Delta,{\bar \Delta})
e^{2\pi i\Delta\tau}e^{-2\pi i{\bar \Delta}{\bar\tau}} .
\label{b4}
\eeq
For a unitary theory, $\rho$ is the number of states with eigenvalues
$L_0 = \Delta$, ${\bar L}_0 = {\bar\Delta}$, as can be seen by inserting
a complete set of states into the trace.  For a nonunitary theory, $\rho$
is the difference between the number of positive-norm and negative-norm
states, although the definition of trace can be changed to make all states
contribute positively.

We can now extract $\rho$ from $Z$ by contour integration.  Treat $\tau$
and $\bar\tau$ as independent complex variables (this is not necessary,
but it simplifies the computation), and let $q=e^{2\pi i\tau}$ and
${\bar q}=e^{2\pi i{\bar\tau}}$, so
\beq
\rho(\Delta,{\bar \Delta}) =
{1\over(2\pi i)^2} \int{dq\over q^{\Delta+1}}
{d{\bar q}\over {\bar q}^{{\bar \Delta}+1}} Z(q,{\bar q}) .
\label{b5}
\eeq
For notational simplicity, I will suppress the $\bar\tau$ dependence, and
restore it only at the end of the computation.  The basic trick is to note
that
\beq
Z(\tau) = e^{{2\pi i c\over24}\tau}Z_0(\tau)
\label{b6}
\eeq
and to use the modular invariance of $Z_0$ to rewrite the contour integral
in a form suitable for a saddle point approximation:
\beq
Z(\tau) = e^{{2\pi i c\over24}\tau}Z_0(-{1/\tau}) =
e^{{2\pi i c\over24}\tau}e^{{2\pi i c\over24}{1\over\tau}}Z(-{1/\tau})
\label{b7}
\eeq
and thus
\beq
\rho(\Delta) = \int d\tau\, e^{-2\pi i\Delta\tau}
e^{{2\pi ic\over24}\tau}e^{{2\pi ic\over24}{1\over\tau}}Z(-{1/\tau}) .
\label{b8}
\eeq

The key to a saddle point approximation is to separate the integrand
into a rapidly varying phase and a slowly varying prefactor.  Let us assume
for the moment---we will have to check this---that $Z(-1/\tau)$ varies
slowly near the extremum of the phase.  For large $\Delta$, the extremum
of the exponent is then
\beq
\tau\approx i\sqrt{c/24\Delta} \ .
\label{b9}
\eeq
Substituting \rref{b9} back into the integral, we obtain
\beq
\rho(\Delta) \approx \exp\left\{2\pi\sqrt{c\Delta\over6}\right\}Z(i\infty) ,
\label{b10}
\eeq
yielding the Cardy formula \rref{a2}.

We must now check the saddle point approximation.  From \rref{b4},
\beq
Z(i/\epsilon) = \sum \rho(\Delta)e^{-2\pi \Delta/\epsilon}
\label{b11}
\eeq
If the lowest eigenvalue of $L_0$ is $\Delta_0=0$, then $Z(i/\epsilon)$
approaches a constant as $\epsilon\rightarrow0$, and the saddle point
approximation is good.  But if $\Delta_0\ne0$, the factor $Z(-1/\tau)$
in \rref{b8} varies rapidly near the putative saddle point, and the
approximation is not valid.  This is easily corrected, however.  Define
\beq
{\tilde Z}(\tau) = \sum \rho(\Delta)e^{2\pi i(\Delta-\Delta_0)\tau} =
e^{-2\pi i\Delta_0\tau} Z(\tau) ,
\label{b12}
\eeq
which goes to a constant as $\tau\rightarrow i\infty$.  Then the integral
for $\rho$ becomes
\beq
\rho(\Delta) = \int d\tau e^{-2\pi i\Delta\tau}
e^{-2\pi i\Delta_0{1\over\tau}}
e^{{2\pi ic\over24}\tau}e^{{2\pi ic\over24}{1\over\tau}}
{\tilde Z}(-{1/\tau})
\label{b13}
\eeq
For $\Delta$ large, this integral {\it can\/} be evaluated in a saddle
point approximation, giving
\beq
\rho(\Delta) \approx
\exp\left\{2\pi\sqrt{(c-24\Delta_0)\Delta\over6}\right\}\rho(\Delta_0) =
\exp\left\{2\pi\sqrt{c_{\hbox{\scriptsize\it eff}}\Delta\over6}\right\}
\rho(\Delta_0) .
\label{b14}
\eeq
Equation \rref{b14} is the generalization of \rref{a2} to theories in
which $\Delta_0\ne0$.

At first sight, the assumption that $\Delta_0=0$ seems innocuous.  But
there is a well known conformal field theory for which this assumption
fails, Liouville theory.  The Liouville action contains a single (albeit
interacting) scalar field, and canonical quantization gives standard
creation and annihilation operators \cite{Thorn}.  The density of states
should thus behave like that of an ordinary scalar field: we should use
equation \rref{a2} with $c=1$.  On the other hand, the central charge
$c_{\scriptstyle\mathit Liou}$ in Liouville theory is determined by
the coupling constants, and can chosen arbitrarily, so the naive Cardy
formula can give an arbitrarily large density of states.  The solution,
as noted by Kutasov and Seiberg \cite{Kutasov}, is that the minimum value
of $L_0$ is not zero for normalizable states in Liouville theory.  Instead
\cite{Seiberg},
\beq
\Delta_0 = {c_{\scriptstyle\mathit Liou}-1\over24}
\label{b15}
\eeq
and thus $c_{\hbox{\scriptsize\it eff}} = c_{\scriptstyle\mathit Liou}
- 24\Delta_0  = 1$ in \rref{b14}, as expected from canonical quantization.
This example is directly relevant to the BTZ black hole, since (2+1)-%
dimensional gravity induces a Liouville theory at spatial infinity
\cite{Carlip4,Henneaux2}, and the central charge \rref{a1} can be
understood as arising from this Liouville theory.

Another example of an ``effective central charge'' will be useful later.
Start with standard affine Lie algebra\footnote{I use the metric $g_{ab}
=\Tr\,T_aT_b$ to raise and lower indices; this convention leads to an
occasional factor of two difference with some expressions in the literature.
In \rref{b18}, $Q$ is defined by $f_{abc}f^{abd} = Q\delta_c^d$.}
\beq
[J^a_m,J^b_n] = if^{ab}{}_cJ^c_{m+n} + kmg^{ab}\delta_{m+n,0}
\label{b17}
\eeq
with the usual affine Sugawara construction for the Virasoro generators,
\beq
L_n = {1\over2k+Q}\sum_p g_{ab}:J^a_p J^b_{n-p}: \ ,
\label{b18}
\eeq
which satisfy the algebra \rref{b2}.  This theory has a central charge
$c$ determined by the group, and its asymptotic density of states
is given by equation \rref{a2}.  Now consider the deformed Virasoro
algebra \cite{Freericks,Sakai} generated by
\beq
{\tilde L}_n
  = L_n + in\alpha_aJ^a_n + {k\over2}\alpha_a\alpha^a\delta_{n0} .
\label{b19}
\eeq
It is easy to check that the ${\tilde L}_n$ again satisfy the Virasoro
algebra \rref{b2}, but with a new central charge
\beq
{\tilde c} = c + 12k\alpha_a\alpha^a .
\label{b20}
\eeq
But the redefinition \rref{b19} has not changed the Hilbert space, so
the asymptotic behavior of the density of states should not be
affected.

Again, the answer lies in the failure of the naive Cardy formula.  The
shift of $L_0$ means that the lowest eigenvalue of ${\tilde L}_0$ is
no longer zero, but rather
\beq
{\tilde \Delta}_0 = {k\over2}\alpha_a\alpha^a .
\label{b20a}
\eeq
The effective central charge in \rref{b14} is thus
$c_{\hbox{\scriptsize\it eff}} = {\tilde c} - 12k\alpha_a\alpha^a = c$,
and the deformation \rref{b19} does not change the asymptotic density
of states.  Like the Liouville case, this example is directly relevant
to the BTZ black hole: in Refs.\ \cite{BBO} and \cite{Banados2},
the central charge \rref{a1} is obtained by precisely such a shift.

It is perhaps worth emphasizing the peculiarity of equation \rref{b14}
for models with $\Delta_0<0$.  A negative value of $\Delta_0$ implies
an {\it increase\/} in the asymptotic density of states: it is as if
a model with $\Delta_0=-1$ had $24$ extra bosonic oscillators.  As we
shall see in section \ref{sec4}, the relative contributions to
$c_{\hbox{\scriptsize\it eff}}$ from $c$ and $\Delta_0$ can depend
on the choice of boundary conditions, and it is possible that the
central charge in Strominger's approach to black hole entropy might
come entirely from a large negative value of $\Delta_0$.  It would be
valuable to understand explicitly---perhaps in a simpler model---exactly
what mechanism is responsible for the contribution of $\Delta_0$ to the
density of states.

\section{Counting States: Combinatorics \label{sec3}}

The preceding section dealt with the indirect counting of states via the
partition function.  In this section, I will discuss a more transparent
counting procedure, based on the combinatorics of creation operators.  Let
us begin with a standard example, a single scalar field, whose creation
and annihilation operators form an affine Lie algebra \rref{b17} for
the group $\hbox{\bf R}$ of real numbers.  We choose a vacuum such that
\beq
J_n |0\rangle = 0 \quad \hbox{for $n>0$}
\label{c1}
\eeq
and create excited states by acting with creation operators $J_{-n}$.
Since $[L_0,J_{-n}] = nJ_{-n}$ and $L_0 |0\rangle = \alpha_0|0\rangle$
for some constant $\alpha_0$,
\beq
L_0 \left( J_{-n_1}J_{-n_2}\dots J_{-n_m}\right) |0\rangle  =
(\alpha_0 + n_1 + n_2 + \dots + n_m) |0\rangle .
\label{c2}
\eeq
The number of states with $L_0 = \Delta$ is thus simply the number of
distinct ways the quantity $\Delta-\alpha_0$ can be written as a sum of
integers.  This is the famous partition function $p(\Delta-\alpha_0)$ of
number theory, whose asymptotic behavior is \cite{Ramanujan}
\beq
\ln p(\Delta-\alpha_0) \sim 2\pi\sqrt{\Delta/6} \ .
\label{c3}
\eeq
This behavior agrees, as it should, with the Cardy formula \rref{a2}
for $c=1$.

Extensions of this result to more than one field appear frequently
in the string theory literature.  However, a more general form seems
not to be widely known \cite{Brigham}.  Suppose we start with bosonic
``creation operators'' $\phi^{(M_n)}_n$, with conformal dimensions
\beq
[L_0,\phi^{(M_n)}_n] = \beta n\phi^{(M_n)}_n .
\label{c4}
\eeq
Here $\beta$ is a constant, and the index $M_n$ distinguishes fields
with identical dimensions.  Let $\gamma(n)$ denote the degeneracy at
conformal dimension $\beta n$, i.e., $M_n=1,\dots,\gamma(n)$.  We allow
$\gamma(n)$ to be zero for some values of $n$---the conformal dimensions
need not be equally spaced.  Now, suppose that the asymptotic behavior
of the sum of degeneracies is
\beq
\sum_{n\le x}\gamma(n) \sim Kx^u
\label{c5}
\eeq
for large $x$.  Then the number of states with $L_0 = \Delta$ can be
shown to grow as
\beq
\ln \rho(\Delta) \sim {1\over u}[u+1]^{u/(u+1)}
\left[ Ku\Gamma(u+2)\zeta(u+1)\right]^{1/(u+1)}
\left[ {\Delta/\beta} \right]^{u/(u+1)} .
\label{c6}
\eeq

For a scalar field, $\gamma(n)=1$ and hence $K=u=1$; it is then easily
checked that \rref{c6} reproduces \rref{c3}.  For $D$ fields, $\gamma(n)=D$,
and $K=D$; the effect is equivalent to the introduction of central charge
$c=D$ in \rref{a2}.  But equation \rref{c6} is considerably more general.
Consider, for example, a set of fields with conformal dimensions $j(j+1)$
and multiplicities $2j+1$, i.e.,
\beq
\gamma(n) = \left\{ \begin{array}{ll}
            2j+1 & \hbox{if $n=j(j+1)$} \\
            0    & \hbox{otherwise} \end{array} \right.
\label{c7}
\eeq
An easy computation shows that $K=u=1$, so the asymptotic density
of states is again given by \rref{c3}, even though the states are no
longer evenly spaced.  I will return to this example in section
\ref{sec6}.

For fermionic creation operators, no corresponding result exists in
the literature, but the generalization of \rref{c6} is straightforward.
The key observation is that the bosonic generating function $p(q)$ and
the fermionic generating function ${\tilde p}(q)$, given by
\begin{eqnarray}
p(q) &=& \prod_{n=1}^\infty (1-q^n)^{-\gamma(n)} = \sum\rho(n)q^n
\nonumber\\
{\tilde p}(q) &=& \prod_{n=1}^\infty (1+q^n)^{\gamma(n)}
  = \sum{\tilde\rho}(n)q^n ,
\label{c8}
\end{eqnarray}
satisfy ${\tilde p}(q)^{-1}p(q) = p(q^2)$, or equivalently,
\beq
\sum_m \rho(m){\tilde\rho}(n-2m) = \rho(n) .
\label{c10}
\eeq
Using this relation and equation \rref{c6}, it is fairly easy to show
that
\beq
\ln {\tilde\rho}(\Delta) \sim {1\over u}[u+1]^{u/(u+1)}
\left[ Ku\Gamma(u+2)\zeta(u+1)(1-2^{-u})\right]^{1/(u+1)}
\left[ {\Delta/\beta} \right]^{u/(u+1)} .
\label{c11}
\eeq
For $u=1$, in particular, the only difference between expressions
\rref{c11} and \rref{c6} is an extra factor of $1/\sqrt2$ in \rref{c11},
corresponding to the well-known fact that a fermionic oscillator
contributes a factor of $1/2$ to the central charge.

Unlike the partition function methods of the preceding section, the
combinatoric techniques described here explicitly display the states
that contribute to the entropy.  Unfortunately, there is a price to pay:
we need to start with a much more concrete description of the vacuum and
the operator content of the theory.  Ultimately, however, some counting
procedure like this will be necessary to complete our understanding of
black hole entropy.

\section{Where Do the Black Hole Degrees of Freedom Live? \label{sec4}}

In typical approaches to black hole statistical mechanics, the degrees
of freedom associated with the entropy are assumed to live on or
near the horizon.  Strominger's derivation, on the other hand, is
based on a central charge \rref{a1} that describes the asymptotic
behavior at spatial infinity.  Yet another suggestion, due to
Martinec \cite{Martinec}, is that the central charge is a result
of ``anomaly inflow''---the entropy comes from D-brane dynamics at the
horizon, but the conformal anomaly is transported to spatial infinity
by the coupling to bulk degrees of freedom.

To understand this issue better, it is helpful to review the derivation
of the central charge in the Chern-Simons formulation of (2+1)-dimensional
gravity.  A careful canonical analysis has been given by Ba{\~n}ados
\cite{Banados2} and by Ba{\~n}ados, Brotz, and Ortiz \cite{BBO} (see
also \cite{Oh}), and I will not repeat it here, but will give a brief
heuristic derivation of their results.

It is well known that diffeomorphisms in a Chern-Simons theory are
equivalent on shell to field-dependent gauge transformations.  Indeed,
the Lie derivative of the gauge potential (or connection one-form) $A$ is
\beq
{\cal L}_\xi A = d(\xi\cdot A) + \xi\cdot dA
  = D_A(\xi\cdot A) + \xi\cdot F
\label{d1}
\eeq
where $D_A$ is the exterior gauge-covariant derivative, $F$ is the
field strength, and the dot denotes contraction of a vector with the
first index of a form.  The Chern-Simons field equations tell us that
$F=0$, and the remaining term in \rref{d1} may be recognized as a
gauge transformation with an infinitesimal parameter $\epsilon^a =
\xi^\mu A_\mu{}^a$.

Now consider a slice at constant time with an $S^1$ boundary, which may
be a black hole horizon or spatial infinity.  Pick a radial coordinate
$\rho$ such that the boundary is located at $\rho=\rho_0$, and choose a
gauge condition
\beq
A_\rho{}^a=\alpha^a
\label{d2}
\eeq
near the boundary, where $\alpha^a$ is a fixed element of the Lie
algebra.  Up to possible quantum corrections, the generator of gauge
transformations at the boundary is \cite{Banados2}
\beq
{\cal G}[\epsilon] = -{k\over2\pi} \int_{S^1}\epsilon_a A_\phi{}^a d\phi,
\label{d3}
\eeq
so by \rref{d1}, the generator of diffeomorphisms is
\beq
{\cal G}[\xi] = -{k\over2\pi} \int_{S^1}
  \left( {1\over2}\xi^\phi g_{ab}A_\phi{}^a A_\phi{}^b
  + \xi^\rho g_{ab}\alpha^a A_\phi{}^b \right) d\phi .
\label{d4}
\eeq
The factor of $1/2$ in the first term reflects the fact that both
copies of $A_\phi$ contribute to the Poisson brackets: schematically,
$\{\xi^\phi (A_\phi)^2, F\} \sim 2\xi^\phi A_\phi\{A_\phi, F\}$.
To preserve the gauge condition \rref{d2}, we should take $\xi^\rho$ and
$\xi^\phi$ to be independent of $\rho$.

Now, a Chern-Simons theory on a manifold with boundary induces a
Wess-Zumino-Witten model on the boundary, with an affine Lie algebra
that is essentially generated by the $A_\phi$.  More precisely, if we
write
\beq
A_\phi{}^a = {1\over k}\sum_{n=-\infty}^\infty J^a_n e^{in\phi} ,
\label{d5}
\eeq
the currents $J^a_n$ will obey the algebra \rref{b17}.  The first term
in \rref{d4} is thus closely related to the Virasoro generator \rref{b18},
and if we can restrict $\xi^\rho$, we have a chance of recovering the
Virasoro algebra \rref{b2}.

In particular, Ba{\~n}ados observed that if we choose $\xi^\rho =
-\partial_\phi\xi^\phi$, we recover an algebra with a central charge
that, up to quantum corrections, is equal to the value \rref{a1}
found by Brown and Henneaux.  This should not be surprising in light of
the model discussed at the end of section \ref{sec2}.  For this choice
of $\xi^\rho$, the generator \rref{d4} becomes
\beq
{\cal G}[\xi] = -{k\over2\pi} \int_{S^1}
  \xi^\phi \left( {1\over2}g_{ab}A_\phi{}^a A_\phi{}^b
  + \alpha_a \partial_\phi A_\phi{}^a \right) d\phi ,
\label{d4a}
\eeq
which leads to precisely the shift \rref{b19} of the Virasoro generators.
Using the results from \ref{appen1} that $\alpha_a\alpha^a=1/2$
and $k=\ell/4G$ for the BTZ black hole, we see that the shift in $c$ is
$3\ell/2G$, in agreement with \rref{a1}.

But it is also clear that the generators \rref{d4} contain many other
Virasoro subalgebras  \cite{BBO}.  If, for example, we choose $\xi^\rho =
-\beta\partial_\phi\xi^\phi$, we obtain a central charge $c(\beta) =
3\beta^2\ell/2G$.  We must somehow determine which choice gives us the
``right'' algebra; that is, we must decide what boundary conditions to
place on the diffeomorphisms.

There are a number of natural choices, which are unfortunately
not all equivalent.  For example, we might fix the induced metric
$g_{\phi\phi}$ on the boundary, by requiring that
\beq
{\cal L}_\xi g_{\phi\phi} = 0 = \xi^\rho\partial_\rho g_{\phi\phi}
  + 2 \partial_\rho\xi^\phi g_{\phi\phi} ,
\label{d6}
\eeq
where $\rho$ is the proper radial coordinate \rref{A5}.  A simple
calculation then shows that for a boundary at Schwarzschild coordinate
$r=r_0$,
\beq
\xi^\rho = - {N(\infty)\over N(r_0)}\partial_\phi\xi^\phi ,
\label{d7}
\eeq
where $N(r)$ is the BTZ lapse function \rref{A2}.  From equation
\rref{b20}, we thus obtain a central charge
\beq
{\tilde c}(r_0) =
c + \left({N(\infty)\over N(r_0)}\right)^2 {3\ell\over2G} .
\label{d9}
\eeq
The naive Cardy formula \rref{a2} would thus give the standard BTZ
entropy \rref{a4} for a boundary at infinity, but a ``blue-shifted''
entropy proportional to $N(\infty)/N(r_0)$ for a boundary at $r=r_0$.
But this is precisely the kind of situation in which \rref{a2} is not
to be trusted, since the minimum eigenvalue $\Delta_0$ of $L_0$ is
also blue-shifted.  Indeed, as we saw at the end of section \ref{sec2},
these shifts cancel in the effective central charge in the generalized
Cardy formula \rref{b14}, and the actual entropy is independent of $r_0$.
Note, though, that we can no longer claim that this entropy is given by
equation \rref{a4}, unless we can control both $c$ and $\Delta_0$ at
the boundary.

Rather than fixing the intrinsic geometry $g_{\phi\phi}$ at the
boundary, we might equally plausibly fix the extrinsic curvature.  For
example, we could fix the radial form of York's ``extrinsic time,''
$\Pi = \sqrt{g_{\phi\phi}}g^{\phi\phi}k_{\phi\phi}$, where
$k_{\phi\phi}$ is the extrinsic curvature of our $S^1$ boundary
viewed as a submanifold of a constant-time slice.  Fixing $\Pi$
requires that
\beq
{\cal L}_\xi \Pi = 0 = \partial_\rho(\xi^\rho\Pi)
  + \partial_\phi(\xi^\phi\Pi) ,
\label{d10}
\eeq
and a straightforward computation shows that
\beq
\xi^\rho = - {N(r_0)\over\partial_\rho N(r_0)}\partial_\phi\xi^\phi .
\label{d11}
\eeq
The corresponding central charge now varies from Strominger's value
\rref{a1} at spatial infinity to {\em zero\/} at the horizon.  Once
again, however, the variation of $\Delta_0$ cancels this effect in the
computation of the entropy.

As yet another alternative, we might choose to fix the mean curvature
$k = g^{\phi\phi}k_{\phi\phi}$ at the boundary.  Since $k$ is a scalar,
this requires that
\beq
{\cal L}_\xi k = 0 = \xi^\rho\partial_\rho k ,
\label{d12}
\eeq
and thus $\xi^\rho=0$.  This choice is physically appealing, since
the condition for an apparent horizon on a time slice of vanishing
mean curvature is that $k=0$ \cite{Steif}.  On the other hand, such
an apparent horizon is equally well determined by the condition that
$\Pi=0$.

We can learn three basic lessons from this analysis:
\begin{enumerate}
\item At least in the Chern-Simons formulation of (2+1)-dimensional
gravity, the central charge of an induced conformal field theory at a
boundary can depend sensitively on the location of the boundary and
the choice of boundary conditions.
\item For the counting of states, this dependence may not matter, since
the {\em effective\/} central charge in the generalized Cardy formula
\rref{b14} may not change.
\item To use \rref{b14} to compute the entropy of the BTZ black hole,
we must control not only the central charge, but also the eigenvalue
$\Delta_0$, on some boundary.
\end{enumerate}

\section{Lowest Virasoro Eigenvalues \label{sec5}}

We have seen that if we wish to use the Cardy formula to compute the BTZ
black hole entropy, we must know both the central charge and the lowest
eigenvalue of $L_0$ on some boundary.  Unfortunately, the general
arguments of Brown and Henneaux \cite{Brown} and Ba{\~n}ados
\cite{Banados2} tell us little about the eigenvalue $\Delta_0$; for
that, we need more information about the relevant conformal field theory.

One way to obtain such information comes from supersymmetry.  As Coussaert
and Henneaux have observed \cite{Coussaert}, the massless BTZ black hole
is supersymmetric, and lies in the Ramond sector of the theory (i.e., the
Killing spinors are periodic).  Similarly, anti-de Sitter space---the
``$M=-1/8G$'' BTZ black hole---is supersymmetric, and lies in the
Neveu-Schwarz sector (the Killing spinors are antiperiodic).  Suppose
the standard superconformal algebra,
\begin{eqnarray}
[L_m, L_n] &=& (m-n)L_{m+n} + {c\over12}m(m^2-1)\delta_{m+n,0} \nonumber\\
\left[L_m, G_n\right] &=& ({1\over2}m-n)G_{m+n} \nonumber\\
\{G_m,G_n\} &=& 2L_{m+n} + {c\over3}(m^2-{1\over4})\delta_{m+n,0} ,
\label{e1}
\end{eqnarray}
applies to the boundary conformal field theory.  In the Ramond sector,
the generators $G_m$ have integer moding, and the lowest possible
eigenvalue of $L_0$ will be
\beq
L_0 |0_{\mathit R}\rangle = {c\over24} |0_{\mathit R}\rangle .
\label{e2}
\eeq
In the Neveu-Schwarz sector, the $G_m$ have half-integer moding, and the
lowest weight is
\beq
L_0 |0_{\mathit{NS}}\rangle = 0 .
\label{e3}
\eeq
The $M=0$ black hole thus has $L_0 = c/24$, and anti-de Sitter space
has $L_0=0$.\footnote{The $L_0$ values in Ref.\ \citen{Strominger}
differ from these by an additive constant of $-c/24$.  This constant
was chosen to adjust $L_0$ to vanish for the $M=0$ black hole.  To use
the Cardy formula for the density of states, however, we must normalize
$L_0$ according to the algebra \rref{e1}.}

If we can consider anti-de Sitter space to be part of our Hilbert
space, and if the canonical analysis of Ba{\~n}ados in Ref.\
\citen{Banados2} can be extended to give the superconformal algebra
\rref{e1}, we can then argue that the lowest eigenvalue of $L_0$ is
in fact $\Delta_0=0$.  If this is the case, the generalized Cardy formula
\rref{b14} reduces to \rref{a2}, and Strominger's analysis gives the
correct black hole entropy.

There is one subtlety in this argument, however.  Although the $M=0$
BTZ black hole and anti-de Sitter space are certainly both supersymmetric,
we do not know a priori {\em which\/} set of Virasoro generators $L_n$
appears in the superconformal algebra \rref{e1}.  As we saw in section
\ref{sec4}, the canonical algebra of boundary diffeomorphisms, at the
horizon or at infinity, contains many copies of the Virasoro algebra
with different central charges, and we do not know which of these should
be associated with anti-de Sitter space.

In particular, the deformation \rref{b19} of the Virasoro algebra of a
WZW model can be extended to the supersymmetric case, with the same shift
in central charge.  In addition to the currents $J^a_n$, a supersymmetric
WZW model contains a set of fermionic oscillators $\psi^a_n$ in the
adjoint representation \cite{Kiritsis}.  It is not hard to check that
the deformation
\beq
{\widetilde G}_n = G_n - 2i\sqrt{k}n\alpha_a\psi^a_n ,
\label{e3a}
\eeq
accompanied by the shift \rref{b19} of the $L_n$, gives a new
superconformal algebra \rref{e1} with the shifted central charge
\rref{b20}.  Note that in the Neveu-Schwarz sector, equation \rref{e3a}
shifts the operators $G_{\pm1/2}$, and the ambiguity can be reformulated
as a question of which of these shifted operators annihilates the anti-de
Sitter vacuum.

Supersymmetric Liouville theory again offers a cautionary tale.
The super-Liouville model is a superconformal field theory, with
an algebra \rref{e1} that can be constructed explicitly from the
fields.  It is tempting to conclude that supersymmetry should
force the minimum eigenvalue of $L_0$ to be zero.  But in fact,
the stress-energy tensor of super-Liouville theory contains an
``improvement'' term of the form \rref{b19}, and  $L_0$ is shifted
by a constant \cite{Liao}
\beq
{\tilde\Delta}_0 = {1\over24}(c - {3\over2}) ,
\label{e3x}
\eeq
yielding an effective central charge of $c_{\hbox{\scriptsize\it eff}}
= 3/2$, as one would expect from counting oscillators.  As in the
nonsupersymmetric version \cite{Seiberg}, the candidate for an
$\hbox{SL}(2,{\bf C})$-invarant vacuum state is not normalizable,
and does not lie in the Hilbert space built from the oscillators of
the model.

Whether the same problem occurs for (2+1)-dimensional gravity can
probably be determined only by a careful extension of a canonical
analysis like that of Ref.\ \citen{Banados2}, with close attention
paid to the relationships between boundary conditions for the
diffeomorphisms and their superpartners.  An important step in this
direction has recently been taken by Ba{\~n}ados et al.\ \cite{BBCHO},
who examine the asymptotic algebra of symmetries in (2+1)-dimensional
supergravity and construct a superconformal algebra.  I believe,
however, that their description of the symmetry algebra is not yet
explicit enough to determine the spectrum, and thus the eigenvalue
${\tilde\Delta}_0$ and the effective central charge.

I will end this section by pointing out a numerical coincidence that
may have a deeper meaning.  The Virasoro generator $L_0$ of equation
\rref{b18} involves an important zero-mode term.  For $\hbox{SL}
(2,{\bf R})$, with the conventions described in \ref{appen1}, this
contribution is
\beq
L_0 = {1\over k-2}\left( -J_0{}^2 + J_1{}^2 + J_2{}^2 \right) +
      \hbox{\it non-zero mode contributions} ,
\label{e4}
\eeq
where $J_0$, $J_1$, and $J_2$ obey the standard $\hbox{SL}(2,{\bf R})$
commutation relations.  From the representation theory of  $\hbox{SL}
(2,{\bf R})$ \cite{Inomata}, we see that for the principle discrete
series,
\beq
L_0 = - {j(j+1)\over k-2} + \hbox{\it non-zero mode contributions} ,
\label{e5}
\eeq
where $j$ is a negative integer or half-integer.  In particular, for
$j=-k/2$,
\beq
L_0 = -{k\over4} + \hbox{\it non-zero mode contributions}
    = -{\tilde \Delta}_0 + \ldots ,
\label{e6}
\eeq
where ${\tilde \Delta}_0$ is the shift in the lowest eigenvalue of $L_0$,
equation \rref{b20a}, for the BTZ black hole.  The deformation \rref{b19}
of $L_0$ thus precisely cancels the zero-mode contribution of the state
with $j=-k/2$.

Now, if the value $j=-k/2$ had been chosen arbitrarily, this would not
be a very significant observation.  But in an $\hbox{SU}(2)$ Chern-Simons
theory, $j=k/2$ is the highest admissible value (the highest integrable
representation), and Hwang has argued that $j=-k/2$ could play an
equivalent role for $\hbox{SL}(2,{\bf R})$ \cite{Hwang,Hwang2}.\footnote%
{See also \cite{Evans}; note that $k$ in that reference is $k/2$ in
the conventions of this paper.} Similarly, in the Euclidean partition
function approach of Ref.\ \citen{Carlip3}, $|j|=|k/2|$ is the maximal
value appearing in the partition function.  This may be accidental, but
it may indicate that the proper deformation \rref{b19} of the Virasoro
algebra simultaneously sets the central charge to the value \rref{a1}
and sets $\Delta_0$ to zero.

\section{Operators and Degrees of Freedom \label{sec6}}

Suppose we can show that the ``correct'' central charge for the
BTZ black hole at some boundary is given by equation \rref{a1}, and
that the corresponding lowest mode of $L_0$ is $\Delta_0=0$, so
Strominger's derivation \rref{a2}--\rref{a4} is correct.  We will still
be left with a question: while the partition function tells us how
many states there are, it does not in itself tell us {\em what\/}
those states are.  It was argued in Ref.\ \cite{Carlip5} (see also
\cite{Balachandran}) that the relevant excitations are ``would-be
gauge'' degrees of freedom, excitations that would normally be pure
gauge, but that become physical as a result of boundary conditions.
But this still does not explicitly express the excitations in terms
of conformal field theory states.  In this final section, I will
speculate briefly on how we might obtain a more transparent description.

A possible starting point is Liouville theory, which provides a well-%
studied example of an ``effective central charge'' of the sort discussed
in section \ref{sec2}.  States in Liouville theory fall into two classes
\cite{Seiberg}: the normalizable ``macroscopic'' (or ``anti-Seiberg'')
states, whose lowest Virasoro eigenvalue is given by \rref{b15}, and the
nonnormalizable ``microscopic'' (or ``Seiberg,'' or ``Hartle-Hawking'')
states, for which $\Delta=0$ can occur.  The division reflects a breakdown
of the usual operator-state correspondence of conformal field theory:
insertions of local operators give nonnormalizable ``microscopic''
states.  This example suggests that if we are looking for BTZ states
with $\Delta_0=0$, we ought to investigate operator insertions rather
than concentrating on the standard (``macroscopic'') Hilbert space
of oscillators.

Now, the fundamental operator in a Wess-Zumino-Witten model is not
the current $J^a$, but the group-valued field $g$.  The conformal
weight of $g$ is not integral: for a spin-$j$ $\hbox{SL}(2,{\bf R})$
representation in the principle discrete series,
\beq
\Delta_j(g) = {j(j-1)\over k-2} ,
\label{f1}
\eeq
where $j$ is a positive integer or half-integer.  The $\hbox{SL}
(2,{\bf R})$ WZW model is not yet understood well enough to determine
which values of $j$ appear in the operator product expansion of $g$,
but let us suppose that all do.\footnote{This is not obvious: for an
$\hbox{SU}(2)$ WZW model, the representations with $j\ge k/2$ completely
decouple.  But the null states for affine $\hbox{SL}(2,{\bf R})$ are
quite different from those for affine $\hbox{SU}(2)$, so the analogy
may be misleading.  As Strominger has pointed out \cite{Strominger3},
one may also worry about whether the operators obtained in this fashion
are mutually local (or, for that matter, whether such a constraint is
necessary).}  We must further determine the multiplicities.  This is
also not known, but a reasonable guess is that the spin $j$ occurs with
a multiplicity $2j+1$, the Plancheral measure for the representation
$j$ \cite{Hwang2}.

(For a compact group $G$, the Peter-Weyl theorem tells us that any
function $F(g)$ can be written as a sum over irreducible representations
of $G$.  The Plancheral formula is, roughly speaking, a generalization
to noncompact groups, allowing $F(g)$ to be expressed as an integral
over irreducible representations.  Let $\widehat G$ be the space of
isomorphism classes of unitary representations of $G$, with $U\in
{\widehat G}$ a representation, and let
\beq
{\widehat U}[F] = \int_G F(h)U(h) dh ,
\label{f2}
\eeq
where $dh$ is the Haar measure.  Then the Plancheral formula tells us
that
\beq
F(g) = \int_{\widehat G} \Tr\left( {\widehat U}(F) U(g)^* \right) d\mu(U) ,
\label{f3}
\eeq
where $d\mu(U)$ is the Plancheral measure \cite{Lang}.  This measure
is thus a reasonable indication of how many times a given irreducible
representation should be counted.)

We can now use equation \rref{c6} to compute the number of states
that can be created by these operators, assuming that there are no
relations among their products.  From the discussion following
equation \rref{c7}, we see that the resulting entropy is
\beq
\ln\rho(\Delta) \sim 2\pi\sqrt{{k\Delta\over3}} .
\label{f4}
\eeq

Equation \rref{f4} {\em almost\/} argrees with Strominger's expression
\rref{a1}--\rref{a2} for the BTZ black hole entropy.  It differs by a
factor of three inside the square root (or a factor of six if we assume
instead that each operator can appear only once and use \rref{c11} for
the density of states).  I do not know how to explain this factor; it
may indicate that this approach to counting states fails.  But it is also
possible that the missing factor reflects the ``bimodular'' properties of
the WZW model.  As Chau and Yamanaka have stressed \cite{Chau}, the
group-valued field $g$ in a WZW model has two independent transformation
properties: it transforms on one side according to the standard
$\hbox{SL}(2,{\bf R})$ Lie algebra, and on the other side under an
appropriate quantum group.  It is plausible that the ``extra'' quantum
group transformation properties lead to a further degeneracy in the number
of states within a given representation of $\hbox{SL}(2,{\bf R})$.

\section{Conclusion}

The derivation of BTZ black hole entropy in Ref.\ \citen{Strominger}
seems too elegant to be wrong.  Unfortunately, it is also too simple to
be completely right: as we have seen---and as Strominger already noted in
\cite{Strominger}---it involves assumptions about the relevant conformal
field theory that are not obviously true for the black hole.  It thus
joins the previous derivations as a highly suggestive, but not quite
complete, computation of black hole entropy from first principles.

With recent developments in higher-dimensional black hole entropy and
anti-de Sitter ``holography,'' the problem of giving a complete,
explicit description of the degrees of freedom responsible for BTZ black
hole entropy seems increasingly urgent.  But the task is perhaps no longer
hopelessly difficult.  I will conclude this paper with a list of open
questions.  The answer to any one of these would represent progress in
our knowledge of BTZ black hole entropy; answers to all would indicate
a fairly solid understanding of the subject.

\begin{enumerate}
\item According to the generalized Cardy formula \rref{b14}, a minimum
Virasoro eigenvalue $\Delta_0<0$ leads to a drastic increase in the
asymptotic growth of the density of states; a value $\Delta_0=-1$ has
the same effect as $24$ bosonic oscillators.  Can this effect be
understood explicitly in terms of a counting argument like those of
section \ref{sec3}?
\item For an $\hbox{SL}(2,{\bf R})$ WZW theory, zero modes can give
negative contributions to $L_0$.  Is there any reason to prefer the
spin $j=-k/2$, which would lead to a correction $-24\Delta_0 = 6k =
3\ell/2G$ in the effective central charge, agreeing with equation
\rref{a1}?
\item Is there any natural way to choose among the boundary conditions
for diffeomorphisms described in section \ref{sec4}?
\item Supersymmetry suggests a minimum Virasoro eigenvalue of $\Delta_0=0$,
but the argument seems to fail for super-Liouville theory, presumably
because the candidate for an $\hbox{SL}(2,{\bf C})$-invariant vacuum is
not a normalizable state.  Does this problem extend to (2+1)-dimensional
anti-de Sitter supergravity?  Alternatively, are there other reasons to
expect that $\Delta_0=0$ for some choice of boundary conditions consistent
with the central charge \rref{a1}?
\item Does the breakdown of the operator-state relationship in Liouville
theory \cite{Seiberg} apply as well to the boundary conformal field theory
induced from (2+1)-dimensional gravity?  If so, is there a way to explicitly
count the nonnormalizable (``microscopic'') states?
\item The operator-counting approach of section \ref{sec6} is suggestive,
but it depends on several uncertain assumptions (appearance of all values
of $j$, multiplicities, independence and consistency of operators) and
misses a factor of three in the final answer.  Is there a way to make
this argument more rigorous, or alternatively to demonstrate that it is
incorrect?
\item The counting method of Ref.\ \citen{Carlip2} differs from others
in several respects.  Some of the differences are discussed below in
appendix \ref{appen2}, but others---most notably involving the role of
zero modes---remain mysterious.  Can the connection between this method
and that of Ref.\ \citen{Strominger} be understood more clearly?
\item Can the Euclidean partition function methods of Refs.\ \citen{Carlip3}
and \citen{BBO} be related to Lorentzian state-counting methods?  How
are the states and zero modes mapped from one signature to the other?
\end{enumerate}

\appendix
\setcounter{footnote}{0}

\section{Metrics, Coordinates, and Conventions \label{appen1}}

The BTZ black hole is a solution to the vacuum Einstein equations
in 2+1 dimensions with a negative cosmological constant $\Lambda =
-1/\ell^2$.  In Schwarzschild-like coordinates, the BTZ metric is
\cite{BTZ}
\beq
ds^2 = -N^2dt^2 + N^{-2}dr^2
  + r^2\left( d\phi + N^\phi dt\right)^2
\label{A1}
\eeq
with lapse and shift functions
\beq
N  = \left( -8GM + {r^2\over\ell^2} + {16G^2J^2\over r^2} \right)^{1/2} ,
  \quad N^\phi = - {4GJ\over r^2} \qquad  (|J|\le M\ell) .
\label{A2}
\eeq
The outer (event) and inner horizons are located at
\beq
r_\pm{}^2={4GM\ell^2}\left \{ 1 \pm
\left [ 1 - \left({J\over M\ell}\right )^2\right ]^{1/2}\right \} ,
\label{A3}
\eeq
i.e.,
\beq
M={r_+{}^2+r_-{}^2\over8G\ell^2}, \quad J={r_+ r_-\over4G\ell} \ .
\label{A4}
\eeq
The radial coordinate $r$ is adapted to the circular symmetry of the
solution, and is characterized by the property that a circle of
constant $r$ has a circumference $2\pi r$.  In the exterior region
$r>r_+$, we can instead choose a proper radial coordinate $\rho$ and
a dimensionless time coordinate $\tau$, defined by \cite{Banados2}
\beq
r^2 = r_+{}^2\cosh^2\rho - r_-{}^2\sinh^2\rho , \quad
\tau = t/\ell .
\label{A5}
\eeq
The metric then becomes
\beq
ds^2 = -\sinh^2\rho(r_+d\tau - r_-d\phi)^2 + \ell^2d\rho^2 +
        \cosh^2\rho(r_-d\tau - r_+d\phi)^2 .
\label{A6}
\eeq

Einstein gravity in 2+1 dimensions with a negative cosmological
constant can be reexpressed as a Chern-Simons theory for the group
$\hbox{SL}(2,{\bf R})\times\hbox{SL}(2,{\bf R})$ \cite{Achucarro,Witten},
with gauge potentials (connection one-forms)
\beq
A^{(\pm)a} = \omega^a \pm {1\over\ell} e^a ,
\label{A7}
\eeq
where $e^a\!=\!e_\mu{}^adx^\mu$ is the triad and $\omega^a\!=\!
{1\over2}\epsilon^{abc}\omega_{\mu bc}dx^\mu$ is the spin connection.
The Einstein-Hilbert action becomes
\beq
I_{\hbox{\scriptsize grav}} =
  I_{\hbox{\scriptsize CS}}[A^{(-)}] -I_{\hbox{\scriptsize CS}}[A^{(+)}] ,
\label{A8}
\eeq
where
\beq
I_{\hbox{\scriptsize CS}} = {k\over4\pi}\int_M\,\Tr
  \left\{ A\wedge dA + {2\over3}A\wedge A\wedge A\right\} ,
\label{A9}
\eeq
is the Chern-Simons action.  The value of the coupling constant $k$
depends on the choice of representation and the definition of the
trace in \rref{A9}.  With the choice \cite{BBO}
\beq
T_0 = {1\over2}\left(\begin{array}{cc} 0 & -1\\1 & 0 \end{array}\right)
, \quad
T_1 = {1\over2}\left(\begin{array}{cc} 1 & 0\\0 & -1 \end{array}\right)
, \quad
T_2 = {1\over2}\left(\begin{array}{cc} 0 & 1\\1 & 0 \end{array}\right) ,
\label{A10}
\eeq
one finds that
\beq
k = {\ell\over4G} .
\label{A11}
\eeq
With these conventions, the metric in the affine Lie algebra \rref{b17}
is $g_{ab} = {1\over2}\eta_{ab}$, and $Q=-4$ in equation \rref{b18}.%
\footnote{I use the conventions of Ref.\ \citen{BBO}: $\eta_{ab} =
{\mathrm{diag}}(-1,1,1)$ and $\epsilon_{012}=1$.}

For the BTZ black hole in the coordinates \rref{A6}, the connection
one-forms are
\begin{eqnarray}
A^{(\pm) 0} &=& \pm {r_+\mp r_-\over\ell}\sinh\rho\, (d\tau\pm d\phi)
\nonumber\\
A^{(\pm) 1} &=& \pm d\rho
\nonumber\\
A^{(\pm) 2} &=& {r_+\mp r_-\over\ell}\cosh\rho\, (d\tau\pm d\phi) .
\label{A12}
\end{eqnarray}
{}From equation \rref{d2}, $\alpha^{(\pm)a} = \pm \delta^a_1$, confirming
that $\alpha_a\alpha^a=1/2$.  The fields \rref{A12} can be converted
by a simple gauge transformation to
\beq
A^{(\pm)} = {r_+\mp r_-\over\ell}T_2\, (d\tau\pm d\phi) ,
\label{A13}
\eeq
from which the holonomies can be read off directly.

\section{Polarizations \label{appen2}}

The entropy calculation of Ref.\ \citen{Carlip2} differs from others
in two striking respects.  First, while most approaches treat
left- and right-movers independently, this reference lumps the left- and
right-moving oscillators together and considers states created by both
from a single vacuum.  Second, the computation involves an integration
over a zero mode ${\bar\omega}$ of the spin connection at the horizon.
These two features are actually closely related: as I shall now show,
both reflect the choice of boundary conditions or ``polarization.''

The partition function for the boundary degrees of freedom can be
obtained as a path integral for (2+1)-dimensional gravity on a solid
torus $M$, with appropriate boundary conditions on the fields at
$\partial M\approx T^2$.  Such a path integral may also be interpreted
as determining a state on $\partial M$, viewed as a function of the
boundary data.  The choice of boundary conditions is thus equivalent
to a choice of polarization, that is, of which phase space variables
to treat as ``positions'' in the argument of the wave function.

In the approaches of Refs.\ \citen{Strominger} and \citen{Carlip3},
the boundary data are spatial components $A^{(\pm)}_\phi$ (or
$A^{(\pm)}_z$) of the connection \rref{A7}.  In particular, the
component $\omega_\phi = (A^{(+)}_\phi + A^{(-)}_\phi)/2$ of the spin
connection is fixed at the boundary.  The corresponding partition
functions are
\beq
Z_A^\pm = \sum\rho(N^\pm)\exp\left\{2\pi i\tau(\Delta^\pm + N^\pm)\right\} ,
\label{B1}
\eeq
where $\Delta^\pm$ are the zero modes of $L_0^\pm$ and
\beq
N^\pm = \sum_{i=1}^3 N_i^\pm
\label{B2}
\eeq
are affine $\hbox{SL}(2,{\bf R})$ number operators.

In Ref.\ \citen{Carlip2}, on the other hand, the boundary is fixed
to be a null surface, with boundary data\footnote{Note that the choice
of group generators in \cite{Carlip2} differs from that used elsewhere
in this paper, and that the coupling constant is therefore renormalized to
${\tilde k} = k/\sqrt{2}$.  The superscript $2$ in this section is a Lie
algebra index.} $e_r{}^2 = r_+/\sqrt{2}$ and $e_v{}^2 = 0$, where
$v$ is a null coordinate.  Now, $e_v{}^2$ and $\omega_\phi{}^2$ are
canonically conjugate, so the partition function in this new polarization
can be obtained from \rref{B1} by a functional Fourier transformation
(see, for example, \cite{Verlinde}):
\beq
Z_e = \int [d\omega]
  \exp\left\{ {2i{\tilde k}\over\pi}
  \int \omega_\phi{}^2 e_v{}^2 d\phi\right\} Z^+_A Z^-_A .
\label{B4}
\eeq
Since we are interested in the boundary condition $e_v{}^2=0$, the
exponential term in \rref{B4} drops out, and since only the zero mode
of $\omega_\phi{}^2$ appears in $Z^\pm_A$, only the integration over
this mode is relevant.  Thus
\beq
Z_e = \sum_{N^+,N^-} \rho(N^+)\rho(N^-) e^{2\pi i\tau(N^+ + N^-)}
  \int d{\bar\omega}\, e^{2\pi i\tau(\Delta^+ + \Delta^-)} .
\label{B5}
\eeq

Now, the conformal field theory of Ref.\ \citen{Carlip2} describes
excitations above a fixed black hole background, unlike that of
\cite{Strominger}, for instance, in which the vacuum is anti-de Sitter
space.  In particular, the black hole mass and angular momentum now
determine the zero modes $\Delta^\pm$, and the physical states are fixed
by the condition that $L_0 = 0$ rather than by the relation \rref{a3}.
The integral over $\bar\omega$ is then exactly that of \cite{Carlip2},
and gives a factor of
$$
\exp\left\{ -2\pi i\tau \left({ 2{\tilde k}^2r_+^2\over\ell^2}\right)
\right\}
$$
in $Z_e$.  If we now let $N^+ + N^- = N$, the sum over $N^-$ in \rref{B5}
may be performed by the method of steepest descents.  In particular, let
$\rho(N^\pm)$ be determined by the counting arguments of section \ref{sec3},
with the three oscillators \rref{B2} in each sector:
\beq
\rho(N^\pm) \sim \exp\left\{ 2\pi\sqrt{N^\pm/2} \right\} .
\label{B3}
\eeq
We then obtain
\beq
\sum_{N^-} \rho(N^-)\rho(N - N^-) \sim
\sum_{N^-} \exp\left\{
      \sqrt{2}\pi \left( \sqrt{N^-} +  \sqrt{N-N^-} \right) \right\} \sim
\exp\left\{ 2\pi\sqrt{N} \right\} .
\label{B6}
\eeq
Combining these results, we find that
\beq
Z_e \sim \sum_N \exp\left\{ 2\pi\sqrt{N} \right\}
  \exp\left\{ 2\pi i\tau \left( N - {2{\tilde k}^2r_+^2\over\ell^2}
  \right)\right\} .
\label{B7}
\eeq
The physical state condition $L_0 = 0$ thus requires that $N =
2{\tilde k}^2r_+^2/\ell^2$, and the density of states in \rref{B7}
reproduces the Bekenstein entropy \rref{a4}.

This derivation highlights another key difference between Ref.\
\citen{Carlip2} and Strominger's approach, the use of the ``naive''
central charge $c\approx3$ in \rref{B3} rather than the much larger
central charge \rref{a1}.  This does not necessarily mean that the
two derivations are incompatible---the change of polarization
described here is, in part, a change of basis, and the counting of
states can appear quite different in different bases.  But it is clear
that the zero modes, which are responsible for the first factor in
\rref{B7}, again play a crucial and rather mysterious role.

\vspace{1.5ex}
\begin{flushleft}
\large\bf Acknowledgements
\end{flushleft}

This paper reflects insights (and occasional arguments) that have
come from conversations with a number of people, including Max
Ba{\~n}ados, Ling-Lie Chau, Marty Halpern, Marc Henneaux, Stephen Hwang,
Miguel Ortiz, and Andy Strominger.  This work was supported in part by
National Science Foundation grant PHY-93-57203 and Department of
Energy grant DE-FG03-91ER40674.


\begin{thebibliography}{99}
\bibitem{BTZ} M.~Ba\~nados, C.~Teitelboim, and J.~Zanelli, Phys.\
 Rev.\ Lett.\ {\bf 69}, 1849 (1992).
\bibitem{BHTZ} M.~Ba\~nados, M.~Henneaux,
 C.~Teitelboim, and J.~Zanelli, Phys.\ Rev.\ {\bf D48}, 1506 (1993).
\bibitem{Carlip1} S.~Carlip, Class.\ Quant.\ Grav.\ {\bf 12}, 2853
 (1995).
\bibitem{Strominger} A.~Strominger, hep-th/9712251, J.~High Energy Phys.\
 {\bf 02}, 009 (1998).
\bibitem{Birmingham} D.~Birmingham, I.~Sachs, and S.~Sen, Phys.\ Lett.\
 {\bf B424}, 275 (1998).
\bibitem{Balasubramanian} V.~Balasubramanian and F.~Larsen, Nucl.\ Phys.\
 {\bf B528}, 229 (1998).
\bibitem{Iofa} M.~Iofa and L.~A.\ Pando Zayas, ``Statistical Entropy of
 Magnetic Black Holes from Near-Horizon Geometry,'' hep-th/9803083.
\bibitem{Alwis} S.~P.\ de Alwis, ``Supergravity, the DBI Action and
 Black Hole Physics,'' hep-th/9804019.
\bibitem{Kaloper} N.~Kaloper, ``Entropy Count for Extremal
 Three-Dimensional Black Strings,'' hep-th/9804062.
\bibitem{Maldacena} J.~Maldacena and A.~Strominger, ``$\hbox{\it AdS}_3$
 Black Holes and a Stringy Exclusion Principle,'' hep-th/9804085.
\bibitem{Lee} H.~W.\ Lee and Y.~S.\ Myung, ``Holographic Connection
 between the BTZ Black Hole and 5D Black Hole,'' hep-th/9804095.
\bibitem{Martinec} E.~J.\ Martinec, ``Matrix Models of {\it AdS} Gravity,''
 hep-th/9804111.
\bibitem{Iofa2} M.~Iofa and L.~A.\ Pando Zayas, ``Statistical Entropy of
 Calabi-Yau Black Holes,'' hep-th/9804129.
\bibitem{Behrndt} K.~Behrndt, I.~Brunner, and I.~Gaida, Phys.\ Lett.\
 {\bf B432}, 310 (1998).
\bibitem{Sachs} I.~Sachs, ``On Universality in Black Hole Thermodynamics,''
 hep-th/9804173.
\bibitem{Teo2} E.~Teo, ``Black Hole Absorption Cross-Sections and the
 anti-de Sitter--Conformal Field Theory Correspondence,'' hep-th/9805014.
\bibitem{Lee2} H.~W.\ Lee, N.~J.\ Kim, and Y.~S.\ Myung, ``Dilaton Test
 of Holography between $\hbox{\it AdS}_3\times S^3$ and 5D Black Hole,''
 hep-th/9805050.
\bibitem{Cvetic} M.~Cveti{\v c} and F.~Larsen, ``Near Horizon Geometry
 of Rotating Black Holes in Five Dimensions,'' hep-th/9805097.
\bibitem{Cvetic2} M.~Cveti{\v c} and F.~Larsen, ``Microstates of
 Four-Dimensional Rotating Black Holes from Near-Horizon Geometry,''
 hep-th/9805146.
\bibitem{Hyun} S.~Hyun, ``U-Duality between Three-Dimensional and Higher
 Dimensional Black Holes,'' hep-th/9704005.
\bibitem{Sfetsos} K.~Sfetsos and K.~Skenderis, Nucl.\ Phys.\ {\bf B517},
 179 (1998).
\bibitem{Teo} E.~Teo, Phys.\ Lett.\ {\bf B430}, 57 (1998).
\bibitem{Achucarro} A.~Ach\'ucarro and P.~K.\ Townsend, Phys.\ Lett.\
 {\bf B180}, 89 (1986).
\bibitem{Witten} E.~Witten, Nucl.\ Phys.\ {\bf B311}, 46 (1988).
\bibitem{Witten2} E.~Witten, Commun.\ Math.\ Phys.\ {\bf 121}, 351 (1989).
\bibitem{EMSS} S.~Elitzur et al., Nucl.\ Phys.\ {\bf B326}, 108 (1989).
\bibitem{Carlip2} S.~Carlip, Phys.\ Rev.\ {\bf D51}, 632 (1995).
\bibitem{Banados1} M.~Ba{\~n}ados and A.~Gomberoff,  Phys.\ Rev.\ {\bf D55},
 6162 (1997).
\bibitem{Strominger2} A.~Strominger and J. Maldacena, gr-qc/9801096,
 J.~High Energy Phys.\ {\bf 02}, 014 (1998).
\bibitem{Carlip3} S.~Carlip, Phys.\ Rev.\ {\bf D55}, 878 (1997).
\bibitem{BBO} M.~Ba{\~n}ados, T.~Brotz, and M.~E.\ Ortiz, ``Boundary
 Dynamics and the Statistical Mechanics of the 2+1 Dimensional Black
 Hole,'' hep-th/9802076.
\bibitem{Brown} J.~D.\ Brown and M.~Henneaux, Commun.\ Math.\ Phys.\
 {\bf 104}, 207 (1986).
\bibitem{Cardy} J.~A.\ Cardy, Nucl.\ Phys.\ {\bf B270}, 186 (1986).
\bibitem{Cardy2} H.~W.~J.\ Bl{\"o}te, J.~A.\ Cardy, and M.~P.\ Nightingale,
 Phys.\ Rev.\ Lett.\ {\bf 56}, 742 (1986).
\bibitem{Banados2} M.~Ba{\~n}ados, Phys.\ Rev.\ {\bf D52}, 5816 (1995).
\bibitem{Giveon} A.~Giveon, D.~Kutasov, and N.~Seiberg, ``Comments on
 String Theory on $\hbox{\it AdS}_3$,'' hep-th/9806194.
\bibitem{Thorn} T.~L.\ Curtright and C.~B.\ Thorn, Phys.\ Rev.\ Lett.\
 {\bf 48}, 1309 (1982); Ann.\ Phys.\ (N.Y.) {\bf 147}, 365 (1983).
\bibitem{Kutasov} D.~Kutasov and N.~Seiberg, Nucl.\ Phys.\ {\bf B358},
 600 (1991).
\bibitem{Seiberg} N.~Seiberg, Prog.\ Theor.\ Phys.\ Suppl.\ {\bf 102},
 319 (1990).
\bibitem{Carlip4} S.~Carlip, Nucl.\ Phys.\ {\bf B362}, 111 (1991).
\bibitem{Henneaux2} O.~Coussaert, M.~Henneaux, and P.~van Driel,
 Class.\ Quant.\ Grav.\ {\bf 12}, 2961 (1995).
\bibitem{Freericks} J.~K.\ Freericks and M.~B.\ Halpern, Ann.~Phys.\
 (N.Y.) {\bf 188}, 258 (1988).
\bibitem{Sakai} N.~Sakai and P.~Suranyi, Nucl.\ Phys.\ {\bf B362}, 655
 (1989).
\bibitem{Ramanujan} S.\ Ramanujan and G.\ H.\ Hardy, Proc.\ London
 Math.\ Soc. (ser.\ 2) {\bf 17}, 75 (1918), reprinted in {\it
 Collected Papers of Srinivase Ramanujan}, edited by G.\ H.\ Hardy et al.
 (Chelsea Publishing Company, NY, 1962).
\bibitem{Brigham} N.~A.\ Brigham, Proc.\ Amer.\ Math.\ Soc.\ {\bf 1},
 182 (1950).
\bibitem{Oh} P.~Oh and M.-I.\ Park, ``Symplectic Reduction and Symmetry
 Algebra in Boundary Chern-Simons Theory,'' hep-th/9805178.
\bibitem{Steif} A.~R.\ Steif, Phys.\ Rev.\ {\bf D53}, 5527 (1996).
\bibitem{Coussaert} O.~Coussaert and M.~Henneaux, Phys.\ Rev.\ Lett.\
 {\bf 72}, 183 (1994).
\bibitem{Kiritsis} E.~B.\ Kiritsis and G.~Siopsis, Phys.\ Lett.\
 {\bf B184}, 353 (1987).
\bibitem{Inomata} See, for example, A.~Inomata, in A.~Inomata, H.~Kuratsuji,
 and C.~C.\ Gerry, {\it Path Integrals and Coherent States of $\hbox{SU}(2)$
 and $\hbox{SU}(1,1)$} (World Scientific, Singapore, 1992).
\bibitem{Liao} H.~C.\ Liao and P.~Mansfield, Nucl.\ Phys.\ {\bf B344},
 696 (1990).
\bibitem{BBCHO} M.~Ba{\~n}ados, K.~Bautier, O.~Coussaert, M.~Henneaux, and
 M.~E.\ Ortiz, ``Anti-de Sitter/CFT Correspondence in Three-Dimensional
 Supergravity,'' hep-th/9805165.
\bibitem{Hwang} S.~Hwang, Nucl.\ Phys.\ {\bf B354}, 100 (1991).
\bibitem{Hwang2} S.~Hwang and P.~Roberts, in {\it Pathways to Fundamental
 Theories}, Proc.\ of the 16th Johns Hopkins Workshop on Current
 Problems in Particle Theory, edited by L.~Brink and R.~Marnelius
 (World Scientific, Singapore, 1993).
\bibitem{Evans} J.~M.\ Evans, M.~R.\ Gaberdiel, and M.~J.\ Perry,
 ``The No-Ghost Theorem for $\hbox{\it AdS}_3$ and the Stringy
 Exclusion Principle,'' hep-th/9806024.
\bibitem{Carlip5} S.~Carlip, in {\it Contrained Dynamics and Quantum
 Gravity 1996}, edited by V.~de~Alfaro et al., Nucl.\ Phys.\ {\bf B}
 (Proc.\ Suppl.) {\bf 57}, 8  (1997).
\bibitem{Balachandran} A.~P.\ Balachandran, L.~Chandar, and A.~Momen,
 Nucl.\ Phys.\ {\bf B461}, 581 (1996).
\bibitem{Strominger3} A.~Strominger, personal communication.
\bibitem{Lang} S.~Lang, ${\mathit SL}_2({\bf\it R})$ (Springer, New
 York, 1985).
\bibitem{Chau} L.-L.\ Chau and I.~Yamanaka, Phys.\ Lett.\ {\bf B369},
 226 (1996).
\bibitem{Verlinde} H.~Verlinde, Nucl.\ Phys.\ {\bf B337}, 652 (1990).

\end{thebibliography}
\end{document}